\documentclass[english,aps,prb,reprint,superscriptaddress,floatfix]{revtex4-1}

\usepackage[normalem]{ulem}
\usepackage{amsmath}
\usepackage{amsthm}
\usepackage{amssymb}
\usepackage{graphicx}
\usepackage{xcolor}
\usepackage{hyperref}
\usepackage{setspace}

\begin{document}

\title{Natural Scales in Geographical Patterns}

\author{Telmo Menezes}

\email{menezes@cmb.hu-berlin.de}

\affiliation{Centre Marc Bloch Berlin e.V., Friedrichstr. 191, 10117 Berlin, Germany}

\author{Camille Roth}

\email{roth@cmb.hu-berlin.de}

\affiliation{Sciences Po, m\'edialab, 84 rue de Grenelle, 75007 Paris, France}

\affiliation{Centre Marc Bloch Berlin e.V., Friedrichstr. 191, 10117 Berlin, Germany}

\affiliation{Centre National de la Recherche Scientifique, France}

\begin{abstract}
Human mobility is known to be distributed across several orders of magnitude of physical distances, which makes it generally difficult to endogenously find or define typical and meaningful scales. Relevant analyses, from movements to geographical partitions, seem to be relative to some ad-hoc scale, or no scale at all. Relying on geotagged data collected from photo-sharing social media, we apply community detection to movement networks constrained by increasing percentiles of the distance distribution. Using a simple parameter-free discontinuity detection algorithm, we discover clear phase transitions in the community partition space. The detection of these phases constitutes the first objective method of characterising endogenous, natural scales of human movement. Our study covers nine regions, ranging from cities to countries of various sizes and a transnational area. For all regions, the number of natural scales is remarkably low (2 or 3). Further, our results hint at scale-related behaviours rather than scale-related users. The partitions of the natural scales allow us to draw discrete multi-scale geographical boundaries, potentially capable of providing key insights in fields such as epidemiology or cultural contagion where the introduction of spatial boundaries is pivotal.
\end{abstract}

\maketitle

\section{Introduction}

Geographical scaling has been at the core of a wealth of studies of human mobility. On one hand, physical distances between connected individuals or between related places have repeatedly been shown to hardly obey any distinctive scale, let alone exhibit distinct phases.
Distance frequencies observed in large geotagged datasets of human behaviour usually follow strongly heterogeneous distributions spanning several orders of magnitude, be it for links based on cell phone movements \cite{gonz-unde,devi-scal} and calls \cite{krin-grav,sobolevsky2013delineating,calabrese2011connected}, social media ``check-ins'' \cite{liu2014uncovering,cho-frie} or postings \cite{beiro2016predicting}, commutes \cite{leno-cros,roth-stru} or taxi rides \cite{lian-unra}, or circulation of artifacts \cite{broc-scal}.
On the other hand, this type of data has more recently been used to uncover geographically consistent areas based on clusters of places, movements or interactions \cite{sobolevsky2013delineating,cranshaw2012livehoods,lengyel-2015geo,liu2014uncovering,ratti2010redrawing,de2011commuter,thiemann2010structure,boy2016study,tool-infe,Bassolas2016} where, in essence, the relevant literature generally proceeds on the assumption that all empirical measurements, irrespective of their diverse spatial scales, should be taken into account to form a single global picture. The choice of the appropriate description scale is left to the beholder: \emph{ex ante}, when gathering data within a given bounding box, and often \emph{ex post}, by focusing on a proper description scale. 
Here, behavioural traces spanning several orders of magnitudes are typically aggregated independently of the physical scale they correspond to; then, geographical areas or patterns are uncovered by community detection algorithms; a final level of description is finally chosen according to some criterion. In practice, these methods generally produce dendrograms defining an embedded series of geographical partitions, where lower-level partitions include higher-level ones in a continuum of increasingly coarse description scales.  An appropriate level of the dendrogram is eventually selected because it either maximises some quantity (typically modularity in network-based methods \cite{thiemann2010structure}), yields a clear-cut dichotomy \cite{sobolevsky2013delineating}, or best matches some a priori known description scale \cite{de2011commuter}.
Results are therefore single-scale rather than scale-free: the aim generally consists in discovering or, rather, recovering \emph{a} gold standard geographic partition of a given area --- such as the partition of Belgium into two linguistic communities \cite{sobolevsky2013delineating}, or the breakdown of administrative regions in Great Britain \cite{ratti2010redrawing}. 
While some studies showed that long- and short-distance connections play distinct roles in defining clusters \cite{thiemann2010structure}, the quest for an ideal set of clusters which have to be discovered once and for all (and possibly aggregated into larger blocks \cite{de2011commuter} until a binary dichotomy is reached \cite{sobolevsky2013delineating}) remains pregnant.

We show here that the choice of observation scales is neither exogenous nor univocal. To this end, we demonstrate that it is possible to endogenously uncover a small number of meaningful description scale ranges from apparently scale-free raw data.  In other words, geographical data on human behaviour encloses several coexisting and natural phases which we recover despite the  absence of scale at the lower level of link distance distributions.

Empirically, we rely on human mobility data stemming from Instagram, an online photo-sharing service targeted at smartphone users. Distributions of link distances between successive user locations are unsurprisingly devoid of a typical scale (see Fig.~S2). Admittedly, this heterogeneity corresponds to an entanglement of a variety of human behaviours, ranging for instance from local commuting to long-range travel. The ubiquitous scale-free distribution of distances certainly aggregates links of diverse nature (and, in turn, depends on diverse geographical patterns, such as borders).  However, we draw markedly distinct conclusions than the current state of the art as regards its significance.
Rather than using all the data and discussing the optimality of a high-level observation scale a posteriori, we work the other way around by a priori relying on link scale to blindly define a series of scale-dependent networks. These networks are based on an increasing link distance threshold and thus configure an increasing movement radius. This then yields a series of geographical partitions from which we derive a small number of remarkably consistent high-level observation scales. This simultaneously defines a small set of relevant low-level scales, in terms of link distances or movement radii. 
 
In other words, we disentangle the endogenous scale structure by exhibiting phase transitions based on the similarity of geographical patterns.
Therefore, while we acknowledge that territories are potentially structured by partially overlapping partitions, we contend that the underlying behaviour corresponds to only a few sensible scales.
 We liken this implicit finding to the explicit, man-made hierarchies which can be found in more traditional top-down approaches relying on discrete ontologies featuring a small number of embedded spatial scales, such as administrative divisions (e.g. NUTS) \cite{de2011commuter}.

\section{Empirical approach} 

We obtained datasets of human movement in a variety of geographical regions from public data extracted from Instagram over a period of $16$ months (See \emph{Materials and Methods} section for a complete and formal description of the techniques and algorithms discussed here). Modern smartphones contain geolocation technology that can be employed to geotag photos, and many users accept this setting. Instagram associates photos with the identifier of the users who took them, and it also timestamps them. 

The global adoption of Instagram in the last few years make it a considerable trove of geosocial data for a nascent literature. Regarding mobility in particular, several recent works demonstrated that reliable mobility data could be extracted from photo-posting and check-in platforms \cite{beiro2016predicting,cho-frie,cranshaw2012livehoods,liu2014uncovering}, while Instagram appears to be both.
Focusing on individual movements, \cite{silva-2013-fsq} shows that geographical patterns inferred from Instagram are similar to those found in data stemming in Foursquare, a so-called ``check-in'' platform where users typically broadcast their position to their friends (however, temporal patterns and posting behaviours differ between the two platforms, suggesting that they correspond to distinct Internet uses).
A more recent study \cite{boy2016study} deals more precisely with sociospatial patterns and divisions at the level of cities.  It describes how Instagram may be used to reveal clusters of users and places which are qualitatively meaningful and consistent with the manual analysis of the information made available by users on their account (describing for instance their profession, affiliations, interests).

For the purpose of this work we are not interested in the photos themselves, but only in the metadata. By tracking the places where a given user took photos we can infer the plausible relatedness between any two given locations in a region.

We focus on nine different areas, that were chosen to offer a diversity of cases according to several criteria: Belgium, Portugal, Poland, Ukraine, Israel, the wider Benelux region and the cities of Berlin and Paris. We have thus five countries of various sizes, a transnational region in Western Europe and two cities.

\subsection{Voronoi diagrams and human movement graphs}

We work with two main formalisms: Voronoi diagrams~\cite{aurenhammer1991voronoi} and weighted graphs of intensity of human movement between places. Given a pre-defined set of geographical positions in a region -- let us call them seeds -- the Voronoi diagram partitions the plane in cells, such that every cell corresponds to a seed. Voronoi diagrams have the property that any point in a cell is closer to the cell seed than to any other seed. This allows us to discretise the space, by assigning each photo in a region to a Voronoi cell. The pre-defined positions are also the vertices in the graph, while the weight of an edge is simply the number of users that the two vertices share --- that is, the number of users that took at least one photo in each location.

From these weighted graphs we generate distributions of edge distances which we divide into $100$ percentiles. This constitutes the full spectrum of scales that we work with. The use of percentiles of the distance distribution allows for the definition of comparable scales across regions, while the absolute distance values that correspond to each scale are determined by the aggregate of human movement in a given region. A graph connecting the positions in a region is generated for each one of these percentiles, by considering only connections with a distance up to the given percentile. In other words, for each region, the graph at scale $s$ gathers $s$\% of all observed movements ranked by increasing distance, \hbox{i.e.} movements up to the distance radius that corresponds to percentile $s$.

\subsection{Geographical clusters and boundaries}

Scale-dependent graphs are partitioned using a \emph{community detection algorithm}. The detection of geographical clusters often relies on network-based community detection method, for instance in \cite{calabrese2011connected} where county-level borders are reconstructed by maximising modularity of graph communities in US-based cell phone data, in \cite{liu2014uncovering} which is based on a weighted network of check-in trajectories, or \cite{de2011commuter} where Louvain is applied on a commuter network. This class of algorithms attempts to find the partition of a graph with the highest modularity, which is the fraction of edges inside the partitions normalised by the expected fraction on a random graph with the same degree distribution. Modularity thus measures the \emph{strength} of a given partition~\cite{newman2006modularity}.

At this stage we switch to Voronoi diagrams. Their cells can now be assigned to geographical clusters according to the results of the above step. To remove residual noise, a smoothing process is applied over the partitioned diagrams. The actual geographical boundaries can now be computed. Two locations are considered neighbours if their respective Voronoi cells share a boundary. Edges shared by two Voronoi cells assigned to different communities are drawn as boundaries.

\section{Results}

\subsection{Phase transitions and natural scales}

After community detection and smoothing for every percentile-scale, we are in a position to analyse the similarity between scales. More specifically, we are interested in seeing if there are well-defined ranges of scales that are sufficiently similar amongst each other and sufficiently distinct between ranges so that we can talk of natural scales, and reduce the $100$ percentiles to a smaller number of scales.

\begin{figure*}
\centering
\includegraphics[width=\linewidth]{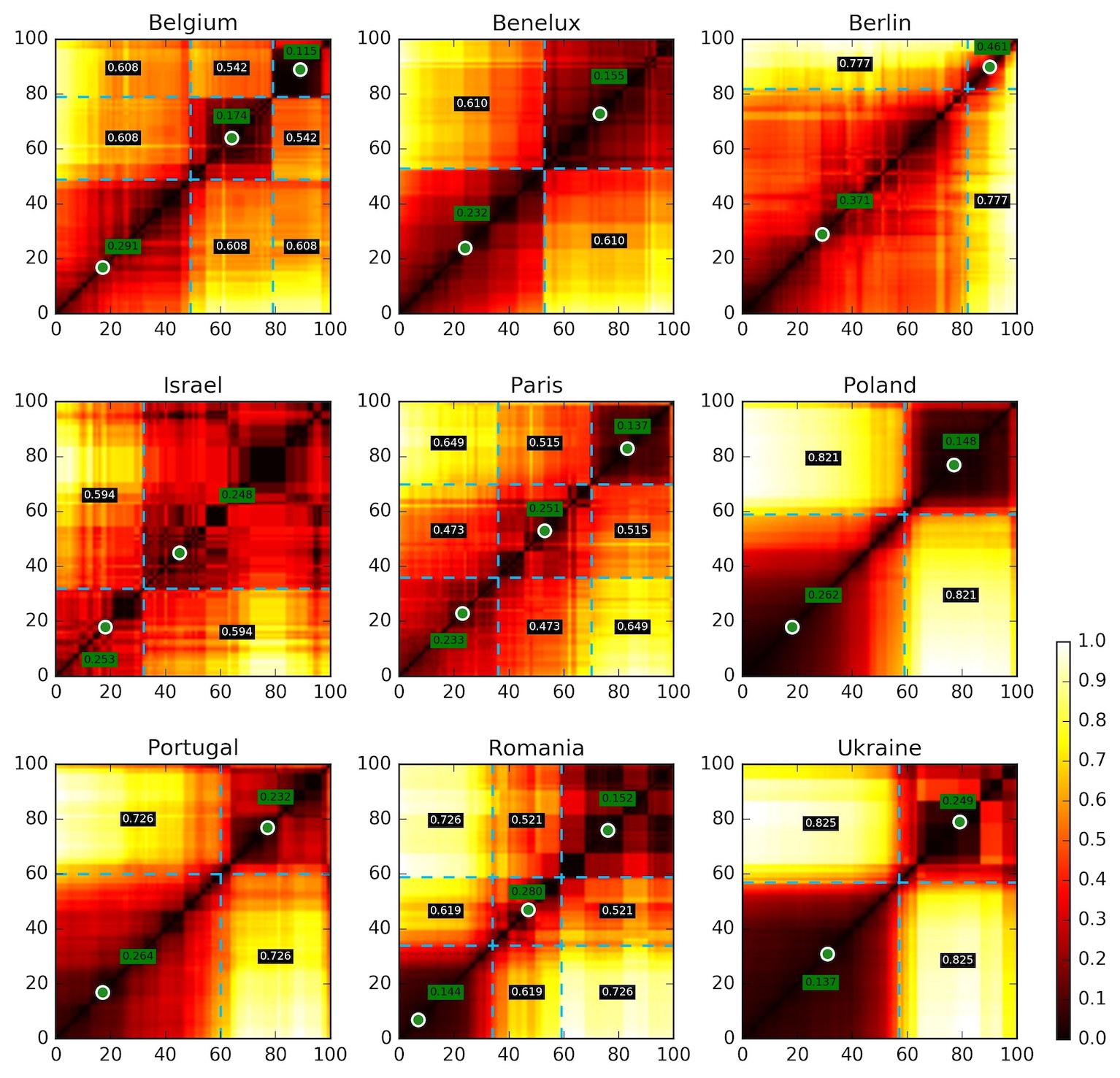} 
\caption{\emph{Scale dissimilarity heat maps.} Dissimilarity values are normalized per region to a [0, 1] scale. Lighter colors represent higher dissimilarity. Pure black (0.0) corresponds to a perfect match, bright yellow (1.0) to the maximum dissimilarity found for the region. Dashed blue lines indicate the discontinuities identified by the breakpoint detection algorithm and, accordingly, natural scales; green dots represent the prototypical scale for each natural scale interval.
The mean absolute dissimilarity value per pair of intervals is shown. A value in a green background corresponds to an internal mean dissimilarity (the interval is being compared to itself); a black background indicates a mean dissimilarity between different intervals.}
\label{heatmaps}
\end{figure*}

\begin{figure*}
\centering
\includegraphics[width=\linewidth]{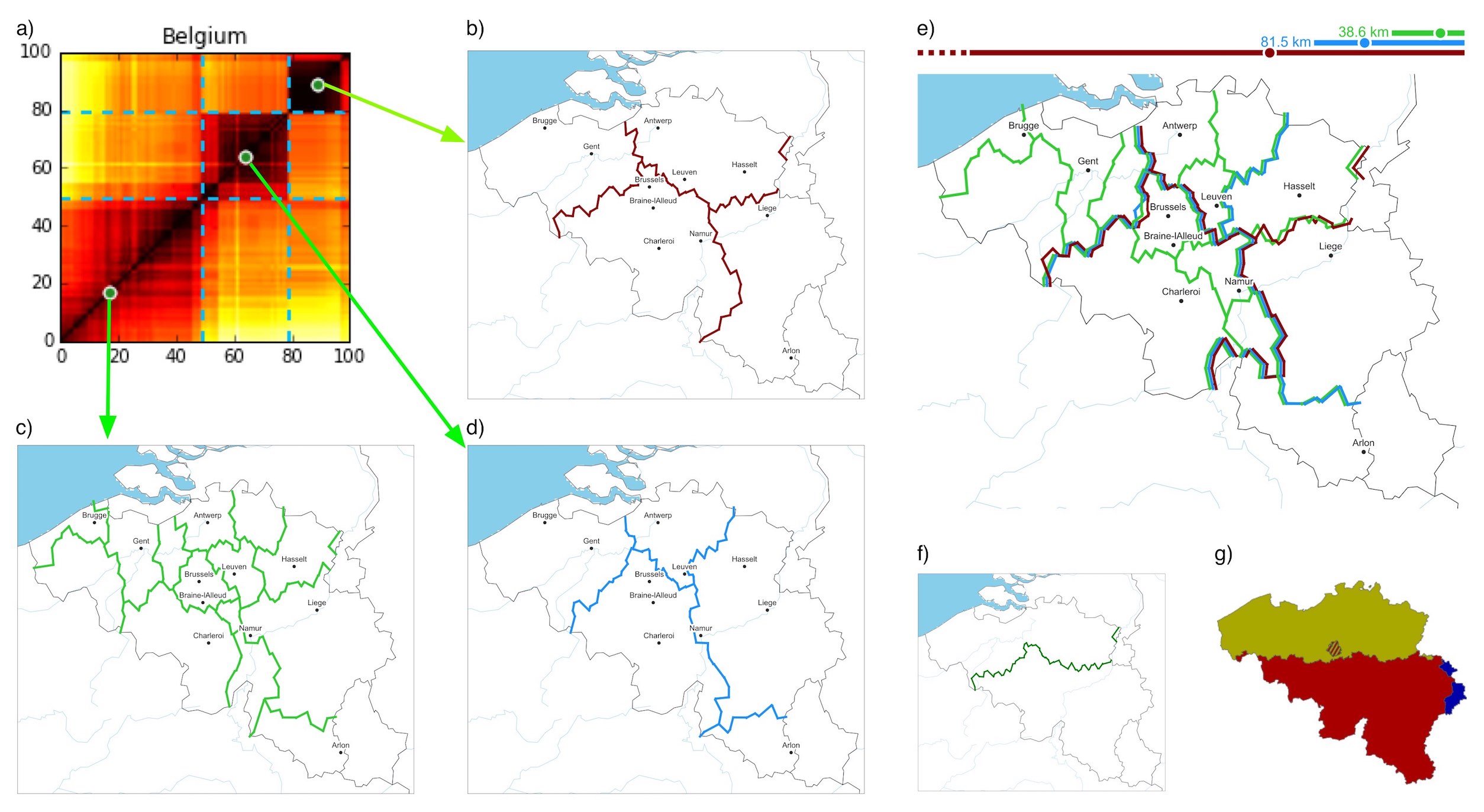} 
\caption{\emph{Belgium borders at different scales.} a) Heat map extracted from figure 1; b) Borders for the long distance scale; c) Borders for the short distance scale; d) Borders for the middle distance scale; e) Multiscale borders; f) Borders based on optimal two community partition of the full graph; g) Language communities of Belgium. All maps except g) were generated by the authors using the Basemap Matplotlib Toolkit ver. 1.0.8 (http://matplotlib.org/basemap/). Map in g) \textcopyright Wikimedia contributers Vascer and Knorck, licensed under CC BY-SA 3.0. The licence terms can be found on the following link:
https://creativecommons.org/licenses/by-sa/3.0}
\label{belgium}
\end{figure*}

Using a simple parameter-free breakpoint detection algorithm we are able to find phase transitions in scale space. We understand ``phase transition'' in a generic way, \hbox{i.e.} it corresponds to an abrupt change in the behaviour of partitions when slightly increasing the movement radius, going from a scale to the next ones. 
For the nine regions under study, our algorithm finds that the scale space is divided into no more than $2$ or $3$ well-differentiated intervals of scales characterised by very similar patterns. Aggregating increasingly long links while remaining below the upper bound of a given interval does not alter significantly the space partition typical of that interval. We call these intervals \emph{natural scales}.
Moreover, the breakpoints automatically found by our algorithm mostly match the visual intuition: in figure~\ref{heatmaps}, we see that these phase transitions are also quite obvious simply by visual inspection.

\subsection{Multi-level partitions and prototypical scales}

Given intervals of similar scales\, or natural scales, it is now desirable to have a method to visualise the boundaries defined by the partitions in those intervals. We propose a simple solution: identify the percentile that best represents the entire interval. We call this prototypical percentile a \emph{prototypical scale} of the region under study. The prototypical scale of a given interval is the percentile of the interval with the corresponding partition that is the most similar to all other partitions in the interval. Prototypical scales found for each region are also represented in figure 1, along with natural scales. By construction, partitions the various scales of a given natural scale should thus roughly resemble the partition of the corresponding prototypical scale. In the following maps, prototypical scales are thus used as visual representations of natural scales.

Figure~\ref{belgium} uses the results for Belgium to illustrate how natural scales correspond to partitions in the map, and how the several natural scales can be combined in a single multiscale map, which provides richer information about the geographical patterns of the region than what is possible with more traditional methods.
By using the full graph (percentile $100$) and forcing the community detection algorithm to find the best partition in two communities, we present a bipartite division of the territory. As can be observed, the resulting partition matches almost perfectly the border between the two largest linguistic communities in Belgium. This is a well known result~\cite{sobolevsky2013delineating, blondel2010regions} and it shows two things. On one hand, when we simplify our method this way, thus making it equivalent to previously published approaches, we obtain similar results, which provides some evidence of correctness. On the other hand, adopting these simplifications is most likely not the best way to unravel the structure of movement patterns in Belgium: to the contrary, for all scale phases, a bipartition does not achieve the best modularity, which usually corresponds to a larger number of geographical areas (Fig. 2).

\begin{figure*}[!th]
\centering
\includegraphics[width=\linewidth]{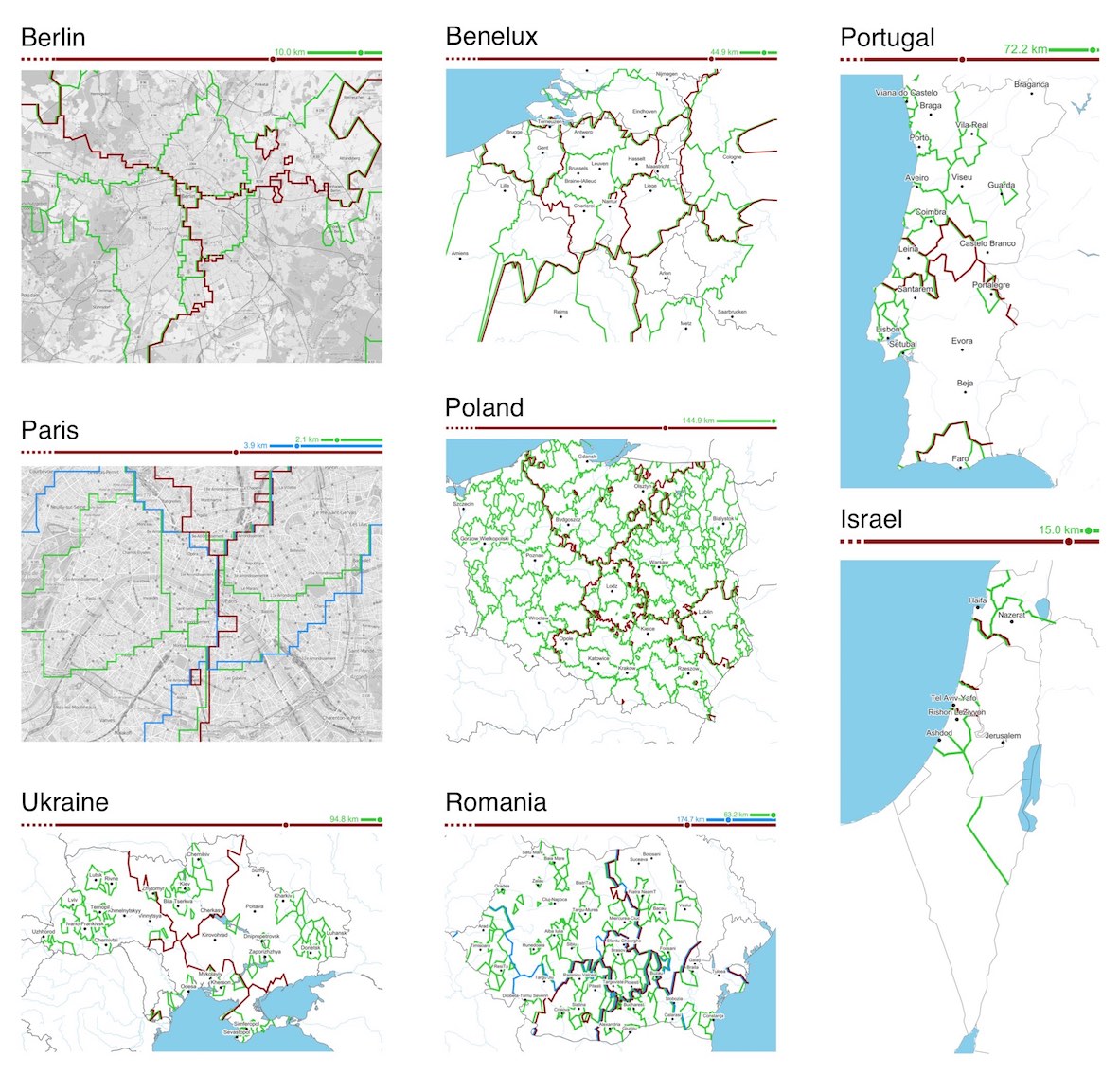} 
\caption{\emph{Several multi-scale maps.} Green corresponds to the smallest natural scale, blue to the middle (if it exists) and red to the largest. All maps were generated by the authors using the Basemap Matplotlib Toolkit ver. 1.0.8 (http://matplotlib.org/basemap/). Map tiles used in the background of the Berlin and Paris maps \textcopyright OpenStreetMap contributors, licensed under CC BY-SA (www.openstreetmap.org/copyright). The licence terms can be found on the following link: http://creativecommons.org/licenses/by-sa/2.0/ .}
\label{maps}
\end{figure*}

\newcommand{\smallbscale}{$38.6$}
\newcommand{\largebscale}{$81.5$}
Following our method, Belgium may be more precisely decomposed as an overlay of three territorial partitions of increasing fineness. The largest natural scale features a partition based on a small number of broad areas whose boundaries correspond to inter-urban mobility, as it only emerges when links longer than \largebscale{}~km  are included.  By contrast, the smallest natural scale is such that most boundaries surround and enclose a local capital; it is based on links smaller than $r\leq$~\smallbscale{}~km. The middle natural scale appears when links between \smallbscale{} and \largebscale{}~km are considered. Interestingly, it diverges from the long-distance scale by only a few boundaries: for instance, while Hasselt is part of a broader Dutch-speaking cluster on the large-scale map, it belongs to the same cluster as Liege at the medium scale.

Finally, in figure~\ref{maps}, we further present the multi-scale maps for all nine regions. 
We also depict the absolute physical distances for all natural scale thresholds. 
Note that the absolute physical meaning of ``large'' or ``small'' scale is heavily region-dependent: for Paris, which is a quite dense metropolis extending over a comparably small area, the smallest natural scale typically covers a range of pedestrian movements ($r\leq2.1$ km). For Berlin, the switch between the small and large natural scales occurs at a radius of $r\leq10.0$ km which could correspond to ``local'' foot, bike or metro trips. For the largest regions such as Poland, Romania or Ukraine, they seem to correspond to a wider range of motorised inter-urban displacements, roughly around the order of magnitude of a hundred of kilometers.

A set of high-resolution maps for all natural scales as well as multi-scale representations of all  regions may be found in the \emph{Supp.~Info.} Performing a thorough socio-geographic analysis of these maps is beyond the scope of this article, but we can identify some features that confirm folk knowledge about certain regions. In Portugal, large scale boundaries delineate the highly touristic beaches of Algarve in the south and fuzzily divide the country into north and south regions, while the short scales provide sensible local partitions, for example the dense city of Oporto and the socio-economic divide between the capital city of Lisbon and the neighbouring but more affluent Cascais/Estoril coastal area. The Benelux map enriches the previous insight on Belgium by providing a broader picture on potential cross-national interfaces -- an achievement not possible with country-specific datasets traditionally used in the literature -- here, the highest scale exhibits a mix of expected international borders (for instance between Belgium and the Netherlands) and fuzzy cross-national spaces (such as the wide commuting area surrounding Luxembourg, or the narrow strips adjacent to the French-Belgian border, \hbox{e.g.} around Lille), while leaving room for cross-border low-scale patterns. Paris features both the traditional east-west sociological partition of the city, while exhibiting more specific activity neighbourhoods at the lower level (Quartier Latin, Belleville, the governmental area).

\begin{figure}[!th]
\centering
\includegraphics[width=\linewidth]{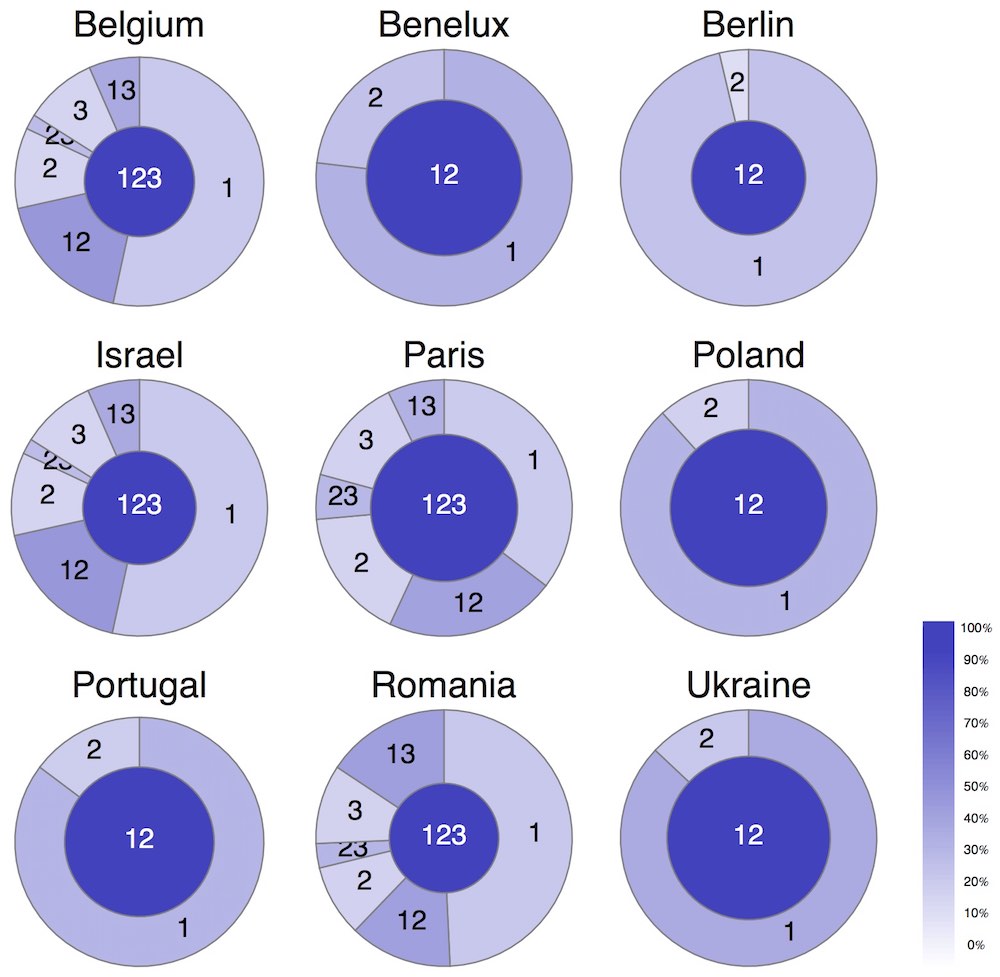} 
\caption{\emph{Fraction of users contributing to each natural scale.} The area of each slice/circle is proportional to the number of users active in the set of scales that it represents (for instance, ``12'' corresponds to users contributing exclusively to scales 1 and 2). User activity is represented by slice darkness, which is proportional to the number of visited locations relative to the maximal activity of a given region (100\%): here, ``123'' users are always the darkest / most active slice, they consistently visit many more locations than other users.}
\label{user-scales}
\end{figure}

\subsection{Scale-dependent user behavior}

Natural scales thus describe geographical areas and boundaries operating within a broad range of scale percentiles, though not beyond. In this respect, they correspond to a discrete spectrum of mobility behaviours which most likely unveil consistent yet distinct spatial practices of the underlying region. How are scales, boundaries and user behaviour related? For one, we observe on figure~\ref{maps} that some regions such as Poland or Romania appear to exhibit a much higher proportion of smaller, lower-scale patterns than other regions such as Paris or Benelux.  We find that these discrepancies have an interpretation in terms of user-level mobility behaviour: regions where movement distance distributions are broadest (i.e. where low and high percentiles correspond to markedly distinct physical distances) also exhibit a much larger amount of small-radius geographical patterns at the shortest scale (see figures~S3 and S4 of \emph{Supp.~Info.}). In other words, we show that the relative amount of patterns across the spectrum of natural scales corresponds to a relative spread of actual physical link distances across that same spectrum.

We further examine the relationship between natural scales and user-level behaviour
by assigning to users the set of natural scales that they contributed to. We consider that users contribute to a natural scale if they perform at least one movement with a distance within the corresponding scale interval. 
Figure~\ref{user-scales} shows a mixed picture.  Overall, the proportion of users contributing exclusively to the highest scales is generally small, while the shortest scales are the most populated. At the same time, the most active users in terms of visited locations (as well as posted photos, see \emph{Supp.~Info.}) are those who span the most scales. From this we conclude that there exists a wide core of users active in all scales, which additionally always gathers a sizable proportion of all users (often the highest proportion). This hints at the fact that natural scales are based on scale-related behaviours rather than scale-related users.

\section{Concluding remarks}

By effectively distinguishing link scales and defining an increasing series of more and more global networks, we show that territories are automatically decomposable into a partially overlapping hierarchy of geographical partitions and, further, that this hierarchy exhibits a remarkably small number of natural scales.
Besides, we fulfilled in the case of spatial mobility networks the ambition of finding natural phases in community partitions based on some notion of resolution (see ref. \cite{traa-sign} for non-geographical scale-free networks). In contrast with the classical expectation that aggregate mobility data is essentially scale-free, we were able to uncover a discrete number of distance thresholds and radii configuring consistent movement patterns.

More broadly, understanding and breaking up mobility patterns as an overlay of a small number of endogenous scale-specific behaviours bears important consequences in diverse fields such as epidemiology, cultural contagion and public policy \cite{ferg-stra,balc-mult,pete-stat} where the low-level modeling of displacements \cite{song-mode,cho-frie,rhee-levy,beiro2016predicting,song-limi} is pivotal: here, the introduction of boundary conditions based on a scaffolding of a small number of natural scales emerging endogenously from the data could prove to be particularly fruitful.

\bigskip\bigskip

\noindent\hfil\rule{0.9\linewidth}{.4pt}\hfil

\bigskip

\section{Methods}

The datasets used in this paper can be obtained by request to the corresponding author. The source code of the programs that implement all the data retrieval and processing tasks discussed in this paper was released as open source and is available in a public repository~\cite{menezes2016}.

\subsection{Data collection and preprocessing}

The Instagram API allows for the collection of all the meta-data and comments of photos on a maximal 5 km radius around a given geographical coordinate. To define points of interest, for countries we use a worldwide database of geographical locations with more than $500$ inhabitants. For cities we simply define a dense enough grid that guarantees that the entire territory is covered, given the 5 km radius around each point. We then query the API for all photo meta-data within the maximal radius around each point. Given the possibility of overlap for locations close to each other, we perform further data processing to remove duplicates and associate each photo with the closest known location.

\subsection{Networks, scales and boundaries}

We use the previous data to generate a weighted graph connecting the set of locations. This graph is undirected and based on user movement, by considering the set of locations where a same user took photos. The weight of an edge represents the number of users who took photos in both the locations connected by the edge. Conventionally, we can represent this graph as $G = \{V, E\}$, where $V$ is the set of locations and $E\subset V\times V\times\mathbb{N}$ the set of weighted edges.

We finally remove all vertices with a very low degree (which we define in \emph{ad hoc} fashion as less than $5$ for all regions). Intuitively, this means that we only consider locations where at least $5$ different users took photos. These very low activity locations are highly susceptible to sampling distortion and introduce noise in the community detection process.

\subsubsection{Percentile graphs}

One graph is generated for each distance percentile, which thereby defines a scale. Let us consider $m_s$ to be the maximum absolute distance for percentile or scale $s$, and $d(e)$ a function that gives the distance between the two vertices of edge $e$. The graph $G_s$ for scale $s$ is then defined as:

\begin{equation}
    G_s = \{V, E_s\}, \text{where } E_s = \{e \in E \,|\, d(e) \le m_s\}
\end{equation}

\subsubsection{Network partitions}

We employ the well-known Louvain method~\cite{blondel2008fast} --- a \emph{de facto} gold standard in network community detection, widely used for the high quality of its results at a low computational cost --- as implemented in the igraph software package~\cite{csardi2006igraph, vincent_traag_2015_35117}. Optimal community detection, like many clustering problems, is probably NP-hard~\cite{fortunato2010community}. The Louvain method is thus an approximation algorithm. It is also non-deterministic. To both achieve higher quality partitions and increase the stability of partitions across scales, we perform $100$ runs of Louvain for each graph, and choose the result that attains the highest modularity. Another common approach is to consider all the outcomes of a large number of runs, and visualise the partitions in a way that assigns visual weights to boundaries in proportion to the number of times they appear~\cite{thiemann2010structure}. Given that we are working with the extra dimension of scale, we avoid this approach for the sake of simplicity.

Notice that community detection is performed on the graph of locations, with no information on the geographic proximity of the vertices. Fortunately, we do find that the communities detected are mostly contiguous, with some noise (see \emph{Supp.~Info.}).

The Louvain method can produce an arbitrary number of partitions. To validate our results, we are also interested in producing bi-partitions. To achieve this, we take the best partition found by Louvain and exhaustively try all possible merges of the given partitions into two. The merge with the highest modularity (although typically lower than the result produced by Louvain) is chosen.

\subsection{Geographical boundary smoothing}

We use this notion of Voronoi neighbourhood to define a smoothing process. From the partition process of the previous section, every location is assigned to a community. If the majority of the neighbours of a location (including the location itself) belong to a different community, then the cell is assigned to this majority community. The process is repeated iteratively, until the previous condition is not triggered.

The geographical boundaries are finally defined by the Voronoi cell boundaries for which the two neighbouring cells' locations do not belong to the same community.

\subsection{Scale similarity, breakpoint detection and natural scales}

\subsubsection{Measuring partition similarity}

Firstly we define a metric of similarity between two partitions of a same set of locations using a Rand index~\cite{rand1971objective}. Consider $V$ the set of locations (as before) and ${P_s}$ and ${P_{s'}}$ two partitions of $V$ produced at scales $s$ and $s'$ by community detection followed by smoothing. A partition is defined as a set of subsets of locations, it is thus included in $\mathcal{P}(V)$.

Let us define a function $\mu^P(i, j)$ that takes the value $1$ if both $i$ and $j$ belong to the same subset of a partition $P$, $0$ otherwise:
\begin{equation}
    \mu^P (i, j) = \left\{\begin{array}{rl}
        1, & \text{if } \exists {X}\in P \text{ \hbox{such that} } i, j\in {X} 
        \\
        0, & \text{otherwise}
        \end{array}\right.
\end{equation}

We can then define the similarity of ${P_s}$ and ${P_{s'}}$ as the ratio between the number of pairs of locations in $V$ that have the same value of $\mu$ for both ${P_s}$ and ${P_{s'}}$ (\hbox{i.e.} they are classified similarly at scales $s$ and $s'$), and the total number of possible location pairs:
\begin{equation}
    \delta(V, {P_s}, {P_{s'}}) = \frac{|\{(i, j)\in V^2, i \ne j, \mu^{{P_s}}(i, j) = \mu^{{P_{s'}}} (i, j)\}|}{\binom{|V|}{2}}
\end{equation}

\subsubsection{Intervals of similar scales}

The above $\delta$ metric allows us to compare graph partitions for each percentile against every other percentile. An immediate application is visual inspection, by generating heatmaps as the ones show in figure 1. A central question to the research being presented in this article is whether partitions ${P_s}$ change smoothly as $s$ increase, or if there are clear discontinuities. The heatmaps indicate quite clearly that the discontinuities do exist.

To identify the breakpoints in partition similarity we introduce another metric, somewhat similar to the concept of modularity in graphs -- albeit even simpler. This metric measures \emph{interval separation}, given a set of breakpoints $B = \{b_0, ..., b_n\}$. Let us also consider the set of intervals defined by these breakpoints: $\mathcal{I}(B) = \{]0..b_0], ]b_0, b_1], ..., ]b_n, 100]\}$. The interval separation for a given $B$ can thus be defined as:

\begin{equation}
    \sigma_B = \frac{\sum_{I \in \mathcal{I}(B)} |I| \cdot \sum_{s,s' \in I} \delta(V, {P_s}, {P_{s'}})}{\displaystyle\max_{b \in B\setminus\{b_0\}} \delta(V, P_{b - 1}, P_b)}
\end{equation}

Intuitively, this is a ratio between the mean similarity within intervals (weighted by the interval size) and the maximum similarity between consecutive partitions in different intervals. The higher the $\sigma$, the greater the similarity between partitions in the same interval compared to the worst case similarity between partitions on both sides of a breakpoint between consecutive intervals.

Using this metric, we define a simple algorithm that iteratively adds breakpoints until $\sigma$ can no longer be improved. We define a minimum interval size of $5$ to avoid isolating noisy outliers. In practice, the minimum interval only has an effect on the two cities, for which the very final scales are indeed quite noisy.

\subsubsection{Prototypical scales}

Intervals thus define natural scales and for a given interval $I$, we define the prototypical scale $s_I$ as the percentile of $I$ with the partition that is the most similar to all other partitions in $I$. To formalise:
\newcommand{\argmax}{\mathrm{argmax}}
\begin{equation}
    s_I = \displaystyle\argmax_{s \in I} \sum_{s' \in I} \delta(V, P_s, P_{s'})
\end{equation}

\subsubsection*{Multi-scale smoothing}

The smoothing method that was previously described can be extended to a set of partitions at $s$ different scales. To each Voronoi tile we assign a tuple consisting of the community number ($c_i$) that the tile $x$ belongs to at each scale after applying Louvain:
\begin{equation}
    t_x = <c_0, ..., c_s>
\end{equation}

Such tuples are treated as values, and the majority rule is applied as before. If a certain tuple is in the majority in the neighbourhood of $t_x$, then $t_x$ takes the value of that entire tuple.

The advantage of this approach is that it leads to a greater overlap of borders from different scales. The cost is that some precision is lost. As can be seen, for example in figure~\ref{belgium}, there are some deviations from the borders at individual natural scales to the borders on the same scales of the multiscale map. We contend that this is a reasonable compromise for the purpose of apprehending the relationship between the different natural scales in a map.

\section*{Acknowledgements}

This paper has been partially supported by grants ``Phantomgrenzen'' and ``Algodiv'' (ANR-15-CE38-0001),  funded respectively by the BMBF (German Federal Ministry for Education and Research) and by the ANR (French National Agency of Research). We warmly thank B\'eatrice von Hirschhausen, Sabine von L\"owis and Karin Casanova for useful discussions and remarks. We are also grateful for constructive anonymous review comments.

\bibliographystyle{plain}
\bibliography{borders}

\end{document}